\documentclass[aps,manuscript,pop,showpacs,preprint,superscriptaddress]{revtex4-1}
\usepackage{lmodern}

\usepackage{lmodern}
\usepackage[T1]{fontenc}
\usepackage{color}
\usepackage{amsmath}
\usepackage{amsthm}
\usepackage{mathdots}
\usepackage{graphicx}
\usepackage{subfigure}
\makeatletter
\theoremstyle{plain}

\ifx\proof\undefined
\newenvironment{proof}[1][\protect\proofname]{\par
	\normalfont\topsep6\p@\@plus6\p@\relax
	\trivlist
	\itemindent\parindent
	\item[\hskip\labelsep\scshape #1]\ignorespaces
}{%
	\endtrivlist\@endpefalse
}
\providecommand{\proofname}{Proof}
\fi

\usepackage{amsthm}\usepackage{latexsym}\usepackage{bm}\usepackage{amsfonts}
\makeatother

\providecommand{\theoremname}{Theorem}

\begin{document}
\title{Canonical Hamiltonian Guiding Center Dynamics and Its Intrinsic Magnetic Moment}

\author{Ruili Zhang}
\affiliation{School of Mathematics and Statistics, Beijing Jiaotong University, Beijing 100044, China}

\author{Jian Liu }
 \thanks{Corresponding author: jliuphy@ustc.edu.cn}
\affiliation{School of Nuclear Science and Technology, University of Science and Technology of China, Hefei 230026, China}
\author{Tong Liu}
\affiliation{School of Mathematics and Statistics, Beijing Jiaotong University, Beijing 100044, China}

\author{Wenxiang Li}
\affiliation{School of Nuclear Science and Technology, University of Science and Technology of China, Hefei 230026, China}

\author{Xiaogang Wang}
\affiliation{Department of Physics, Harbin Institute of Technology, Harbin 150001, China}

\author{Yifa Tang}
\affiliation{LSEC, ICMSEC, Academy of Mathematics and Systems Science, Chinese Academy of Sciences, Beijing 100190, China}

\begin{abstract}
{The concept of guiding center is potent in astrophysics, space plasmas, fusion researches, and arc plasmas to solve the multi-scale dynamics of magnetized plasmas.
In this letter, we rigorously prove that the guiding center dynamics can generally be described as a constrained canonical Hamiltonian system with two constraints in six dimensional phase space, and that the solution flow of the guiding center lies on a canonical symplectic sub-manifold.
The guiding center can thus be modeled as a pseudo-particle with an intrinsic magnetic moment, which properly replaces the charged particle dynamics on time scales larger than the gyro-period.
The complete dynamical behaviors, such as the velocity and force, of the guiding center pseudo-particle can be clearly deduced from the model.
Furthermore, a series of related theories, such as symplectic numerical methods, the canonical gyro-kinetic theory, and canonical particle-in-cell algorithms can be systematically developed based on the canonical guiding center system.
The canonical guiding center theory also provides an enlightenment for the origin of the intrinsic magnetic moment.
}

%\textbf{\normalsize{}Keywords}{\normalsize{}: Guiding center dynamics;
%Constrained Hamiltonian system; Canonical Hamiltonian; Intrinsic Magnetic Moment; Canonical Symplectic Algorithm}{\normalsize\par}
\end{abstract}
\maketitle

Plasmas at different characteristic scales are described by different theories, such as kinetic theory, gyrokinetics, drift kinetics, the two-fluid model, and magnetohydrodynamics (MHD).
The completeness of these hierarchy theories for distinct scale levels is critical to handle complex plasma systems.
The concept of guiding center and gyrokinetic theory have been widely applied to relieve the multi-scale problem arising in magnetized plasmas and have been proven to be powerful in studying physical phenomena above the gyro-motion scale in astrophysics, space physics, fusion research, arc plasmas, etc.
Guiding center dynamics underlies gyrokinetic theory and plays an important role in both fundamental plasma theories and simulation models, achieving fruitful results
\cite{littlejohn1979guiding,littlejohn1983,GCsoliton1991,qin1999gyrokinetic,qin2000gyrocenter,sugama2000gyrokinetic,brizard2007foundations,Song2015,Howes2011,Chang2017,burby2020guiding,Rubin2023,Brizard2023}.
In applications and physical explanations, guiding centers are often conflated with the motions of charged particles.
As an independent hierarchical model, the final link of guiding center dynamics that has yet to be completed is its canonical Hamiltonian theory.
Since the Lagrangian formulation of guiding center was first proposed by Littlejohn in 1983 \cite{littlejohn1983}, researchers have been looking for its canonical Hamiltonian formulation  \cite{littlejohn1979guiding,littlejohn1981hamiltonian,littlejohn1982hamiltonian,tronko2015lagrangian,neishtadt2019hamiltonian}.

Hamiltonian mechanics was introduced by Sir Hamilton in 1833 as a reformulation of Lagrangian mechanics.
It replaces generalized velocities used in Lagrangian mechanics with generalized momenta.
Although both theories provide interpretations of classical mechanics and describe the same physical phenomena, Hamiltonian theory has a closer relationship with geometry, such as symplectic geometry and Poisson structures, and serves as a crucial link between classical and quantum mechanics.
The Hamiltonian formulation of guiding center system forms a classic fundamental topic in the fields of plasma physics and astrophysics \cite{littlejohn1979guiding,littlejohn1983,qin1999gyrokinetic,sugama2000gyrokinetic,cary2009hamiltonian,middelkamp2011guiding,Xiao2020SlowMO}.
.

Meiss and Hazeltine discussed the existence of the canonical coordinates of guiding center systems, but their canonical scheme is not practical in real applications \cite{meiss1990canonical}.
White, Zakharov, and Gao studied the canonical form of guiding center motion in magnetic fields with toroidal flux-surfaces in detail \cite{white2003hamiltonian,gao2012hamiltonian}.
On the other hand, the canonical Hamiltonian formulation can be asymptotically obtained from a noncanonical Hamiltonian form of the guiding center Euler-Lagrange equation  through coordinate transformation based on the Darboux-Lie theorem \cite{zhang2014}.
The main idea of previous works is to find suitable coordinates to express the Hamiltonian function in a 4-dimensional space.
These canonical coordinates thus depend on the specific expression of magnetic field.
In real physical systems, not all magnetic fields have flux coordinates, and it is also challenging to articulate magnetic fields in terms of flux coordinates.
In fact, in most cases we cannot know the specific formulation of the magnetic field in advance.
A general and rigorous canonicalization of guiding center dynamics is therefore very important.

In this letter, we manipulate motion equations of guiding center in the 6-dimensional phase space.
We rigorously prove that the guiding center dynamics can generally be described as a constrained canonical Hamiltonian system with two constraints, and that the solution flow of the guiding center lies on a canonical symplectic sub-manifold.
This new canonicalization scheme is generalized and  does not depend on the property of magnetic fields.
The specific expression and information about magnetic fields are not required in advance.
The new canonical formulation also enhances our physical comprehensions of the guiding center system.
In the 6-dimensional canonical coordinates, a guiding center can thus be modeled as a pseudo-particle with its intrinsic magnetic moment, which properly replaces the charged particle dynamics on time scales larger than the gyro-period.
The complete dynamical behaviors, such as the velocity and force, of the guiding center pseudo-particle can be clearly established from the canonical Hamiltonian, while two constrains play key roles.
The velocity of guiding center exhibits clear origins.
Its parallel velocity is constrained to the direction of the magnetic field by the two dynamical constraints.
And its vertical velocity, which corresponds to the drift motion of guiding center, arises from the coupling of its intrinsic magnetic moment and the  constraints in Lagrangian multipliers.
As for the force acting on guiding center pseudo-particles, the two constraints supply an additional force to counteract the Larmor gyromotion.
Furthermore, a series of related theories, such as symplectic numerical methods, the canonical gyro-kinetic theory, and canonical particle-in-cell algorithms can be systematically developed based on the canonical guiding center system.
When extracting guiding center dynamics from charged particle dynamics in finer space-time scale, the intrinsic magnetic moment of guiding center pseudo-particle emerges naturally.
This cross-scale treatment of guiding center system could shed light on the origin of the intrinsic magnetic moment of other more physically real particles, whereas in quantum physics magnetic moment and spin are considered fundamental intrinsic properties of particles and attributed to their internal symmetries.

We start from Littlejohn's guiding center Lagrangian in the 8D coordinates $(\boldsymbol{X},\dot{\boldsymbol{X}},u,\dot{u})$, i.e.,
\begin{equation}
L(\boldsymbol{X},\dot{\boldsymbol{X}},u,\dot{u})=(\boldsymbol{A}(\boldsymbol{X})+u\boldsymbol{b}(\boldsymbol{X}))\cdot\dot{\boldsymbol{X}}-\left[\dfrac{1}{2}u^{2}+\mu B(\boldsymbol{X})+\Phi(\boldsymbol{X})\right],\label{eq:ol}
\end{equation}
where $\boldsymbol{X}$ denotes the position of the guiding center, $\boldsymbol{A}(\boldsymbol{X})$ denotes the vector potential at $\boldsymbol{X}$, $B(\boldsymbol{X})=||\boldsymbol{B}(\boldsymbol{X})||=||\nabla\times\boldsymbol{A}(\boldsymbol{X})||$ denotes the magnetic field, $\boldsymbol{b}(\boldsymbol{X})=\dfrac{\boldsymbol{B}(\boldsymbol{X})}{B(\boldsymbol{X})}=(b_{1},b_{2},b_{3})^{T}$ is the unit vector along the magnetic field, and $\Phi(\boldsymbol{X})$ denotes the scalar potential.
Here $u$ has a ambiguous physical meaning, because it appears in guiding center Lagrangian but is usually interpreted as the parallel velocity of the original charged particle with respect to the magnetic field.
Similarly, $\mu$ denotes the magnetic moment of the original particle.
The mass and charge of the original particle are normalized to unity.
The Euler-Lagrange equation concludes $\boldsymbol{b}\cdot\dot{\boldsymbol{X}}=u$.
To clarify the physical model and explore the mathematical structure, we replace the variable $u$ in Eq.\,\eqref{eq:ol} by $\boldsymbol{b}\cdot\dot{\boldsymbol{X}}$.
Because $\mu$ is an adiabatic invariant and comes from gyro-symmetry, we then define it as the intrinsic magnetic moment of the guiding center pseudo-particle, and its meaning is no longer related to the original particle.
Now the Lagrangian of guiding center can be written in a purely 6D guiding center coordinates $(\boldsymbol{X},\dot{\boldsymbol{X}})$, i.e.,
\begin{equation}
L(\boldsymbol{X},\dot{\boldsymbol{X}})=\dfrac{1}{2}\dot{\boldsymbol{X}}^{T}M(\boldsymbol{X})\dot{\boldsymbol{X}}+\boldsymbol{A}(\boldsymbol{X})\dot{\boldsymbol{X}}-\left[\mu B(\boldsymbol{X})+\Phi(\boldsymbol{X})\right],\label{eq:DL}
\end{equation}
where the matrix $M(\boldsymbol{X})$ is
\[
M(\boldsymbol{X})=\left(\begin{array}{ccc}
b_{1}^{2} & b_{1}b_{2} & b_{1}b_{3}\\
b_{1}b_{2} & b_{2}^{2} & b_{2}b_{3}\\
b_{1}b_{3} & b_{2}b_{3} & b_{3}^{2}
\end{array}\right).
\]
Supposing $\boldsymbol{B}(\boldsymbol{X})\neq 0$ for magnetized plasmas, the matrix $M(\boldsymbol{q})$ is singular with a constant rank $Rank(M(\boldsymbol{X}))=1$, and satisfies $M^{2}(\boldsymbol{X})=M(\boldsymbol{X})$ and $M^{T}(\boldsymbol{X})=M(\boldsymbol{X})$.\textcolor{black}{{}
The new Lagrangian in Eq.\,\eqref{eq:DL} is still degenerate.
Compared with Littlejohn's Lagrangian, it depends on fewer variables and is more similar to the Lagrangian of charged particle.}
The Hamiltonian formulation can be obtained from the Lagrangian function through Legendre transformation.
If the Lagrangian is non-degenerate, i.e. $\dfrac{\partial^{2}L}{\partial\dot{\boldsymbol{X}}\partial\dot{\boldsymbol{X}}}$ is nonsingular, a canonical Hamiltonian system can be deduced using the standard Legendre map $Leg(\boldsymbol{X},\dot{\boldsymbol{X}})=(\boldsymbol{X},\boldsymbol{p}),$ where the canonical momentum is defined by $\boldsymbol{p}=\dfrac{\partial L}{\partial\dot{\boldsymbol{X}}}=(p_{1},p_{2},p_{3})^{T}.$
The corresponding Hamiltonian function is defined as $H(\boldsymbol{X},\boldsymbol{p})=\boldsymbol{p}^{T}\dot{\boldsymbol{X}}-L(\boldsymbol{X},\dot{\boldsymbol{X}}(\boldsymbol{X},\boldsymbol{p})).\label{eq:Ha0}
$
For degenerate Lagrangian, generalized Hamiltonian theory has also been well developed by Dirac \cite{dirac1950generalized} and A. Van der Schaft \cite{van1987equations}.
When we apply the generalized Hamiltonian theory to the guiding center Lagrangian in Eq.\,\eqref{eq:DL}, the generalized Legendre
transformation leads to
\begin{equation}
\boldsymbol{p}=\dfrac{\partial L}{\partial\dot{\boldsymbol{X}}}=M(\boldsymbol{X})\dot{\boldsymbol{X}}+\boldsymbol{A}(\boldsymbol{X})\Leftrightarrow\begin{alignedat}{1}p_{1} & =b_{1}\boldsymbol{b}\cdot\dot{\boldsymbol{X}}+A_{1}(\boldsymbol{X}),\\
p_{2} & =b_{2}\boldsymbol{b}\cdot\dot{\boldsymbol{X}}+A_{2}(\boldsymbol{X}),\\
p_{3} & =b_{3}\boldsymbol{b}\cdot\dot{\boldsymbol{X}}+A_{3}(\boldsymbol{X}).
\end{alignedat}
\label{eq:pc}
\end{equation}
It outputs $3$ dependent generalized momentum variables $p_{1},\thinspace p_{2}$ and $p_{3}$, which are connected by $3-1=2$ primary constraints \cite{van1987equations}.
From Eq.\,\eqref{eq:pc}, we conclude that $\dfrac{p_{i}-A_{i}(\boldsymbol{X})}{b_{i}(\boldsymbol{q})}$ are equal for $i=1,2,3$, i.e.,
$
\boldsymbol{b}\cdot\dot{\boldsymbol{X}}=\dfrac{p_{1}-A_{1}(\boldsymbol{X})}{b_{1}(\boldsymbol{X})}=\dfrac{p_{2}-A_{2}(\boldsymbol{X})}{b_{2}(\boldsymbol{X})}=\dfrac{p_{3}-A_{3}(\boldsymbol{X})}{b_{3}(\boldsymbol{X})}.
$
The two primary constraints are then expressed explicitly as
\begin{equation}
\begin{alignedat}{1}g^{1}(\boldsymbol{X},\boldsymbol{p}) & =b_{1}(\boldsymbol{X})\left(p_{2}-A_{2}(\boldsymbol{X})\right)-b_{2}(\boldsymbol{X})\left(p_{1}-A_{1}(\boldsymbol{X})\right)=0,\\
g^{2}(\boldsymbol{X},\boldsymbol{p}) & =b_{1}(\boldsymbol{X})\left(p_{3}-A_{3}(\boldsymbol{X})\right)-b_{3}(\boldsymbol{X})\left(p_{1}-A_{1}(\boldsymbol{X})\right)=0.
\end{alignedat}
\label{eq:c3}
\end{equation}
These two primary constraints force $\boldsymbol{p}-\boldsymbol{A}(\boldsymbol{X})$ to follow the direction of magnetic field.
They are steady motion constraints that depend on both position $\boldsymbol{X}$ and canonical momentum $\boldsymbol{p}$, different from other more regular constrained Hamiltonian systems.
The corresponding Hamiltonian of this non-degenerate system is
\begin{equation}
\begin{alignedat}{1}H(\boldsymbol{X},\boldsymbol{p}) & =\boldsymbol{p}^{T}\dot{\boldsymbol{X}}-L(\boldsymbol{X},\dot{\boldsymbol{X}}(\boldsymbol{X},\boldsymbol{p}))\\
 & =\dfrac{1}{2}\dot{\boldsymbol{X}}^{T}M(\boldsymbol{X})\dot{\boldsymbol{X}}+\mu B(\boldsymbol{X})+\Phi(\boldsymbol{X})\\
 & =\dfrac{1}{2}(\boldsymbol{b}\cdot\dot{\boldsymbol{X}})^{2}+\mu B(\boldsymbol{X})+\Phi(\boldsymbol{X}).
\end{alignedat}
\label{eq:H0}
\end{equation}
We further express $(\boldsymbol{b}\cdot\dot{\boldsymbol{X}})^{2}$ in Eq.\,\eqref{eq:H0} using coordinates $(\boldsymbol{X},\boldsymbol{p})$.
Because $\boldsymbol{b}(\boldsymbol{X})$ is a unit vector, i.e., $\sum_{i=1}^{3}b_{i}^{2}=1$, and $b_{i}(\boldsymbol{b}\cdot\dot{\boldsymbol{X}})=p_{i}-A_{i}(\boldsymbol{X})$
in Eq.\,\eqref{eq:pc}, we can write $(\boldsymbol{b}\cdot\dot{\boldsymbol{X}})^{2}$
in the form of $(\boldsymbol{b}\cdot\dot{\boldsymbol{X}})^{2}=\sum_{i=1}^{3}b_{i}^{2}(\boldsymbol{b}\cdot\dot{\boldsymbol{X}})^{2}=\sum_{i=1}^{3}(p_{i}-A_{i}(\boldsymbol{X}))^{2}.$
The Hamiltonian is finally written as
\begin{equation}
\begin{alignedat}{1}H(\boldsymbol{X},\boldsymbol{p}) & =\dfrac{1}{2}\sum_{i=1}^{3}(p_{i}-A_{i}(\boldsymbol{X}))^{2}+\Phi(\boldsymbol{X})+\mu B(\boldsymbol{X}).\end{alignedat}
\label{eq:H1}
\end{equation}
It consists of two parts corresponding to the motion of guiding center and the intrinsic magnetic moment, respectively.
The canonical Hamiltonian in Eq.\,\eqref{eq:H1} together with the two primary constraints in Eq.\,\eqref{eq:c3} determines the complete guiding-center dynamics.
Calculated according to \cite{van1987equations}, the dynamical equations of guiding-center as a constrained Hamiltonian system with two constraints are
\begin{equation}
\begin{alignedat}{1}
\begin{cases}\dot{\boldsymbol{y}} & =J\nabla H(\boldsymbol{y})+\sum_{k=1}^{2}\lambda^{k}J\nabla_{\boldsymbol{y}}g^{k}(\boldsymbol{y}),\thinspace(\boldsymbol{y}=(\boldsymbol{X}^{T},\boldsymbol{p}^{T})^{T}\in R^{6}),\\
0 & =g^{k}(\boldsymbol{y}),\thinspace(k=1,2),
\end{cases}
\end{alignedat}
\label{eq:Hc}
\end{equation}
where $J=\left(\begin{array}{cc}
0 & I_{3}\\
-I_{3} & 0
\end{array}\right)$ is the standard symplectic matrix, and $\lambda^{k} (k=1,2)$ denotes two Lagrangian multipliers.
The canonical guiding center Hamiltonian system has two complex constraints depending on both $\boldsymbol{X}$ and $\boldsymbol{p}$.
We can prove that the solution flow of guiding center lies on a symplectic submanifold and it conserves its energy $H(\boldsymbol{y})$ exactly (see the Supplemental Material)\cite{zhang2014,feng1986,feng1995collected,GNT}.
We now successfully find canonical coordinates for guiding center in the 6-dimensional phase space explicitly.
Unlike previous canonicalization in 4-dimensional space, this canonical scheme does not depend on the specific magnetic field.
The guiding center motion has 4 local degrees of freedom, since it's on a sub-manifold in 6-dimensional space with two constraints.
But the global dimension of guiding center dynamics depends on the property of magnetic fields.
According to Eq.\,\eqref{eq:Hc}, the motion of the guiding center can be regarded as a constrained pseudo-particle with its intrinsic magnetic moment in 6-dimensional phase space.
The primary constraints $g^{k} (k=1,2)$ must be preserved along the solution and satisfy $\dot{g^{k}}=\left(\nabla_{\boldsymbol{y}}g^{k}\right)^{T}\dot{\boldsymbol{y}}=0 \thinspace(k=1,2)$, which are also known as the Casimir functions and explicitly written as
\begin{equation}
\begin{alignedat}{1}\left(g_{\boldsymbol{X}}^{1}\right)^{T}H_{\boldsymbol{p}}-\left(g_{\boldsymbol{p}}^{1}\right)^{T}H_{\boldsymbol{X}}+\lambda^{2}\left\{ g^{1},g^{2}\right\}  & =0,\\
\left(g_{\boldsymbol{X}}^{2}\right)^{T}H_{\boldsymbol{p}}-\left(g_{\boldsymbol{p}}^{2}\right)^{T}H_{\boldsymbol{X}}+\lambda^{1}\left\{ g^{2},g^{1}\right\}  & =0,
\end{alignedat}
\label{eq:sc}
\end{equation}
where $\{\cdot,\cdot\}$ is the Poisson bracket defined by $\{f,l\}=\sum_{i=1}^{3}\left(\dfrac{\partial f}{\partial\boldsymbol{X}_{i}}\dfrac{\partial l}{\partial\boldsymbol{p}_{i}}-\dfrac{\partial f}{\partial\boldsymbol{p}_{i}}\dfrac{\partial l}{\partial\boldsymbol{X}_{i}}\right).$
Without loss of generality, $\left\{ g^{1},g^{2}\right\} \neq0$ holds.
The Lagrangian multipliers can then be calculated from Eq.\,\eqref{eq:sc} as

\begin{equation}
\begin{alignedat}{1}&\lambda^{1}(\boldsymbol{X},\boldsymbol{p})  =\dfrac{\left\{ g^{2},H\right\} }{\left\{ g^{1},g^{2}\right\} }\\
&=\dfrac{(p_3-A_3)\nabla b_1\cdot (\boldsymbol{p}-\boldsymbol{A})-(p_1-A_1)\nabla b_3\cdot (\boldsymbol{p}-\boldsymbol{A})+\mu (\boldsymbol{b}\times \nabla B)_2+(\boldsymbol{b}\times \nabla \Phi)_2}{b_{1}[(\boldsymbol{p}-\boldsymbol{A})\cdot \nabla\times \boldsymbol{b} +\boldsymbol{b}\cdot(\nabla\times \boldsymbol{A})]},\\
&\lambda^{2}(\boldsymbol{X},\boldsymbol{p})  =-\dfrac{\left\{ g^{1},H\right\} }{\left\{ g^{1},g^{2}\right\} }\\
&=\dfrac{-(p_2-A_2)\nabla b_1\cdot (\boldsymbol{p}-\boldsymbol{A})+(p_1-A_1)\nabla b_2\cdot (\boldsymbol{p}-\boldsymbol{A})+\mu (\boldsymbol{b}\times \nabla B)_3+(\boldsymbol{b}\times \nabla \Phi)_3}{b_{1}[(\boldsymbol{p}-\boldsymbol{A})\cdot \nabla\times \boldsymbol{b} +\boldsymbol{b}\cdot(\nabla\times \boldsymbol{A})]}.
\end{alignedat}
\label{eq:Lambda}
\end{equation}

\begin{figure}
\includegraphics[scale=0.6]{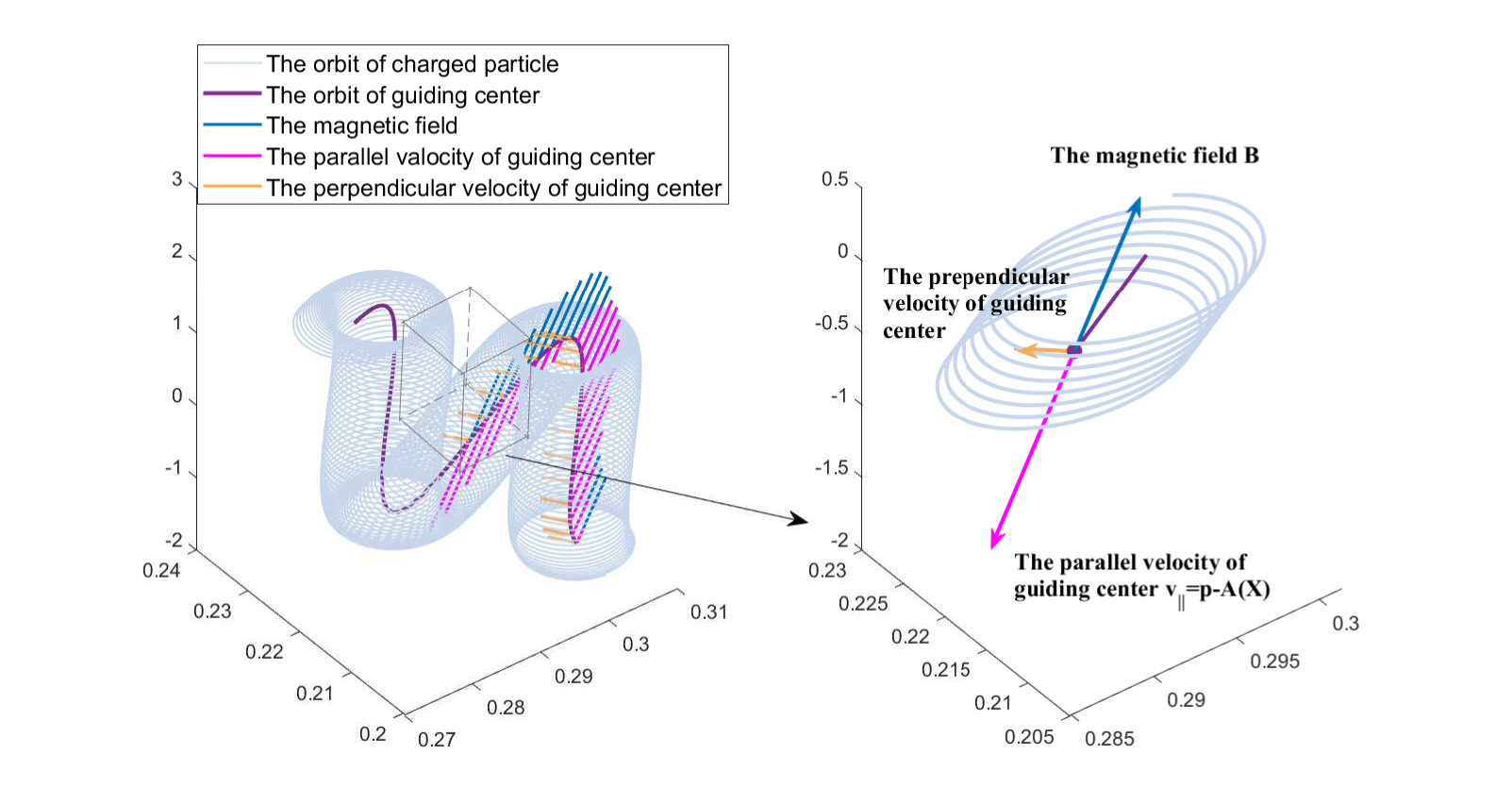}
\caption{The orbits of a guiding center pseudo-particle (in purple) in an external magnetic field, with corresponding charged particle (in grey) for reference. The vectors of magnetic field (in blue), $\boldsymbol{p}-\boldsymbol{A}(\boldsymbol{X})$ (in pink), and perpendicular velocity of the guiding center (in orange) are plotted at different positions along the trajectory. A section of the orbits is zoomed in and shown in the right panel. The initial values of guiding center are $\boldsymbol{X}_0=(0.301,0.207,-1.4)$ and $\boldsymbol{p}_0=(-10.35,15.05,0.2650)$, with the initial values of corresponding charged particle are $\boldsymbol{x}_0=(0.3,0.2,-1.4)$ and $\boldsymbol{p}_0=(-10.7,15.08,0.265)$.}
\label{fig1}
\end{figure}

With the explicit expression of Lagrangian multipliers above, the derivative of the constraints vanish correspondingly, and the global constraints can be replaced by the constraints at the initial point.
Substituting the expressions of $\lambda^{k}(\boldsymbol{y})$ into Eq.\,\eqref{eq:Hc} and releasing the global constraints, we obtain an equivalent ordinary differential equation
\begin{equation}
\begin{cases}
\begin{array}{c}
\dot{\boldsymbol{X}}=\dfrac{\partial H}{\partial \boldsymbol{p}}+\sum_{k=1}^{2}\lambda^{k}(\boldsymbol{X},\boldsymbol{p})\nabla_p g^{k}(\boldsymbol{X},\boldsymbol{p}),\\
\dot{\boldsymbol{p}}=-\dfrac{\partial H}{\partial \boldsymbol{X}}-\sum_{k=1}^{2}\lambda^{k}(\boldsymbol{X},\boldsymbol{p})\nabla_X g^{k}(\boldsymbol{X},\boldsymbol{p}),\\
g^{k}(\boldsymbol{y}_{0})=0,k=1,2,
\end{array}\end{cases}\label{eq:neweq0}
\end{equation}
which can explicitly provide the guiding center dynamical equations as
\begin{equation}
\begin{cases}
\begin{alignedat}{1}
\dot{\boldsymbol{X}}&=\boldsymbol{p}-\boldsymbol{A}(\boldsymbol{X})+\dfrac{(\boldsymbol{p}-\boldsymbol{A})\times \nabla \boldsymbol{b}\cdot (\boldsymbol{p}-\boldsymbol{A})+\mu \boldsymbol{b}\times \nabla B+\boldsymbol{b}\times \nabla \Phi}{(\boldsymbol{p}-\boldsymbol{A})\cdot \nabla\times \boldsymbol{b} +B},\\
\dot{\boldsymbol{p}}&=-\mu \nabla B-\nabla\Phi+(\dfrac{\partial \boldsymbol{A}}{\partial \boldsymbol{X}})^T\dot{\boldsymbol{X}}\\
&+(\dfrac{\partial \boldsymbol{b}}{\partial \boldsymbol{X}})^T\dfrac{(p_1-A_1)/b_1[(\boldsymbol{p}-\boldsymbol{A})\times \nabla \boldsymbol{b}\cdot (\boldsymbol{p}-\boldsymbol{A})]+(\boldsymbol{p}-\boldsymbol{A})\times (\mu\nabla B+\nabla \Phi)}{(\boldsymbol{p}-\boldsymbol{A})\cdot \nabla\times \boldsymbol{b} +B},\\
&g^{k}(\boldsymbol{X}_{0},\boldsymbol{p}_{0})=0,k=1,2.
\end{alignedat}\end{cases}\label{eq:neweq}
\end{equation}
This formulation based on canonical coordinates $(\boldsymbol{X},\boldsymbol{p})$ offer a complete and self-sufficient description of guiding center dynamics, which is distinct from the Littlejohn's  interpretation as the approximation of charged particle dynamics.
The first equation of Eq.\,\eqref{eq:neweq} determines the velocity of pseudo-particle, depicted in Fig.\ref{fig1}.
The parallel velocity $V_{||}=p-A(\boldsymbol{X})$ is constrained in the direction of magnetic field by two primary constraints, where $p$ is now the parallel canonical momentum of the guiding center.
The velocity component perpendicular to the magnetic field takes a more complex form as
$$V_{\perp}=\sum_{k=1}^{2}\lambda^{k}(\boldsymbol{X},\boldsymbol{p})\nabla_p g^{k}(\boldsymbol{X},\boldsymbol{p})=\dfrac{(\boldsymbol{p}-\boldsymbol{A})\times \nabla \boldsymbol{b}\cdot (\boldsymbol{p}-\boldsymbol{A})+\mu \boldsymbol{b}\times \nabla B+\boldsymbol{b}\times \nabla \Phi}{(\boldsymbol{p}-\boldsymbol{A})\cdot \nabla\times \boldsymbol{b} +B}.$$
Through the Lagrangian multipliers $\lambda^{k}(\boldsymbol{X},\boldsymbol{p})$, the intrinsic magnetic moment and two primary constraints interfere the vertical velocity and cause the drift motion of guiding center.
In the perpendicular velocity, $\dfrac{(\boldsymbol{p}-\boldsymbol{A})\times \nabla \boldsymbol{b}\cdot (\boldsymbol{p}-\boldsymbol{A})}{(\boldsymbol{p}-\boldsymbol{A})\cdot \nabla\times \boldsymbol{b} +B}$ corresponds to the curvature drift, $\dfrac{\mu \boldsymbol{b}\times \nabla B}{(\boldsymbol{p}-\boldsymbol{A})\cdot \nabla\times \boldsymbol{b} +B}$ corresponds to the gradient drift, and $\dfrac{\boldsymbol{b}\times \nabla \Phi}{(\boldsymbol{p}-\boldsymbol{A})\cdot \nabla\times \boldsymbol{b} +B}$ corresponds to the electric drift.

\begin{figure}
\includegraphics[scale=0.45]{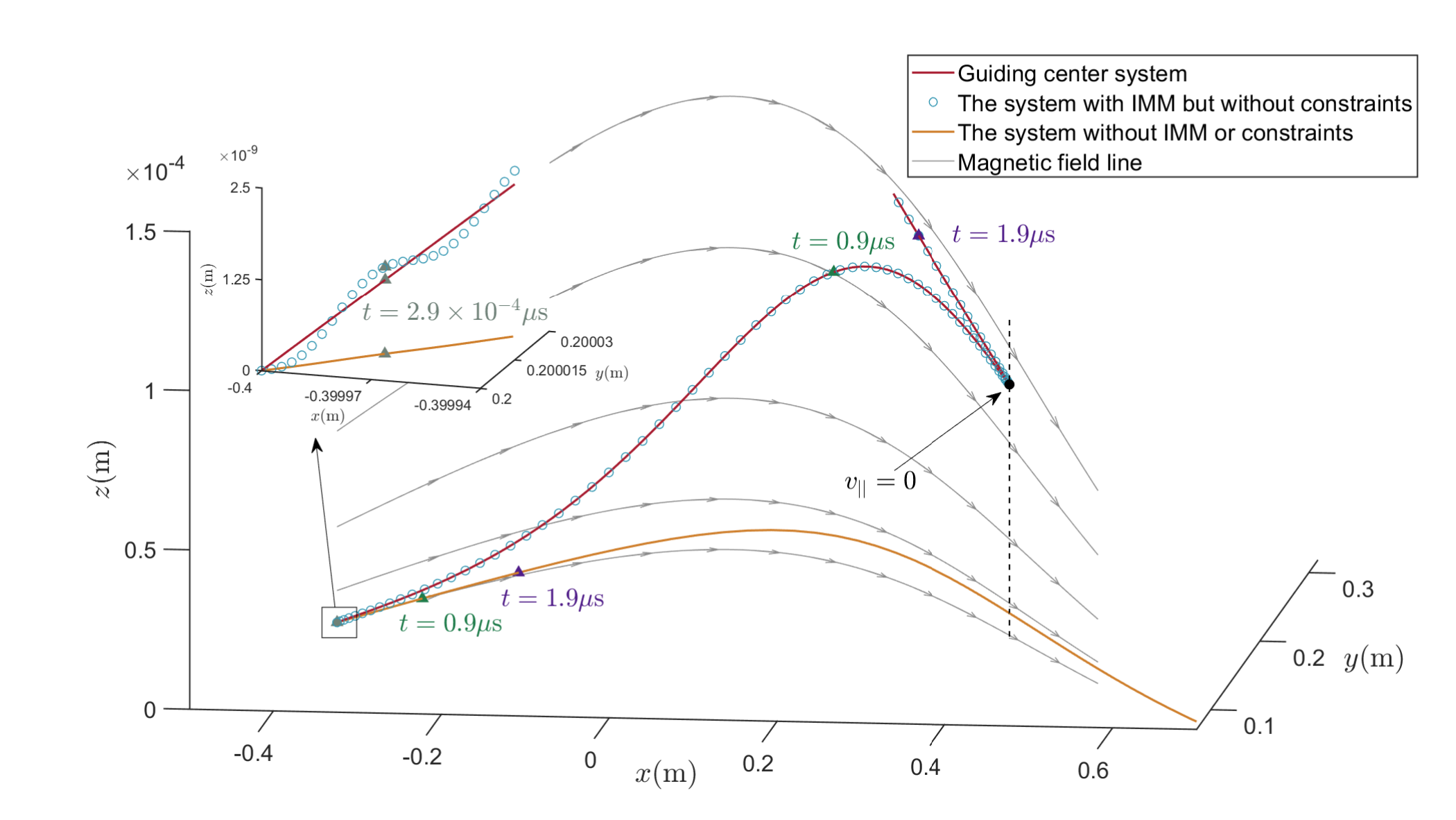}
\caption{The trajectories of three different systems in a magnetic mirror field for comparison.
The initial position of all the three systems is set to be $\boldsymbol{X}_0=(-0.4, 0.2, 0)$ m.
The initial parallel velocity is $V_{||0}= 10^{6}$ $\mathrm{m}/ \mathrm{s}$.  The magnetic moment is $\mu = 4.4458 \times 10^{-18}$.
The red curve depicts the trajectory of the canonical guiding center system.
The orange curve depicts the trajectory of the system without IMM or constraints.
And the light blue circles trace the orbit of the system with the complete Hamiltonian but without constraints.
The triangles with different colors expose the spots of the three systems at different times, respectively.
The zoomed-in window exhibits details of the fast timescale dynamics of the system with IMM but without constraints.
}
\label{fig2}
\end{figure}

The second equation of Eq.\,\eqref{eq:neweq} establishes the force acting on the pseudo-particle
\begin{equation}
\begin{array}{cc}
\dot{\boldsymbol{V}}
=\dot{\boldsymbol{p}}-(\dfrac{\partial A}{\partial \boldsymbol{X}})\dot{\boldsymbol{X}}+\dot{V_{\perp}}=-\mu \nabla B(\boldsymbol{X})-\nabla\Phi(\boldsymbol{X})+\dot{\boldsymbol{X}}\times \boldsymbol{B}(\boldsymbol{X})\\+[-\lambda_1(p_2-A_2)\nabla b_{1}+\lambda_1(p_1-A_1)\nabla b_{2}-\lambda_2(p_3-A_3)\nabla b_{1}+\lambda_2(p_1-A_1)\nabla b_{3}+\dot{V_{\perp}}],\\
\end{array}
\end{equation}
where, $-\mu \nabla B$ is the force related to the intrinsic magnetic moment, $-\nabla\Phi$ denotes the electrostatic force, $\dot{\boldsymbol{X}}\times \boldsymbol{B}(\boldsymbol{X})$ corresponds the magnetic force.
The last term is the force caused by the two constraints, which exactly counteracts the Larmor gyromotion, as shown in Fig.\ref{fig2}.
The third equation in Eq.(\ref{eq:neweq}) determines the initial values of the guiding center, where the initial value of the position $\boldsymbol{X}_{0}$ and the initial value of canonical momentum $\boldsymbol{p}_{0}$ have to satisfy the primary constraints Eq.\,\eqref{eq:pc}.
The initial canonical momentum of the guiding center pseudo-particle is defined as $\boldsymbol{p}_{0}=(V_{||0}+A(\boldsymbol{X}_{0}))\boldsymbol{b}(\boldsymbol{X}_{0})$, where $V_{||0}$ is its initial parallel velocity.

So far, we have provided a general canonicalization scheme for the guiding center system without specific information of magnetic field.
The guiding center canonical coordinates $(\boldsymbol{X},\boldsymbol{p})$ has been explicitly written out in a 6-dimensional phase space.
The guiding center canonical Hamiltonian system is completely described by Eq.\,\eqref{eq:H1} together with the two primary constraints in Eq.\,\eqref{eq:c3}.
The guiding center is then modeled as a pseudo-particle independent of the original charged particle with an intrinsic magnetic moment, and its motion is restrained by two primary constraints on a submanifold.
Its complete dynamics is described by the dynamical equation in Eq.\,\eqref{eq:neweq}.
In canonical coordinates, the symplectic structure of the guiding center system is clear, and symplectic algorithms for guiding center are ready to be constructed.
The canonical gyrokinetic equation for guiding center pseudo-particle distribution function $F(\boldsymbol{X},\boldsymbol{p},t)$ without considering collisional effects can also be easily obtained, which is
\begin{equation}
\frac{\partial F}{\partial t}+\dot{\boldsymbol{X}}\cdot\frac{\partial F}{\partial \boldsymbol{X}}+\dot{\boldsymbol{p}}\cdot\frac{\partial F}{\partial \boldsymbol{p}}=0,
\label{eq:cgkinetic}
\end{equation}
where $\dot{\boldsymbol{X}}$ and $\dot{\boldsymbol{p}}$ satisfy Eq.(\ref{eq:neweq}).

To intuitively display and verify the canonical theory for guiding center pseudo-particle and its effectiveness in applications, we look into some physical systems based on it.
Firstly, we numerically calculate the motion of a guiding center pseudo-particle and corresponding charged particle for reference, using the mid-point rule, in a given magnetic field
\[
\boldsymbol{B}(\boldsymbol{x})=(y,-x,100),
\]
with the corresponding electromagnetic potentials being $\boldsymbol{A}(\boldsymbol{X})  =(-50y,50x,\dfrac{x^{2}+y^{2}}{\text{\ensuremath{2}}})$ and $\Phi(\boldsymbol{X})=\frac{1}{2}x^2+\frac{1}{2}y^2+\frac{1}{2}z^2 $.
Figure\,\ref{fig1} compares the orbits of the guiding center governed by Eq.\,\eqref{eq:neweq} and its original charged particle, respectively.
The trajectory of the guiding center perfectly depicts the motion of the charged particle on time scales larger than the gyroperiod.
In Fig.\,\ref{fig1}, it is demonstrated that $\boldsymbol{p}-\boldsymbol{A}(\boldsymbol{X})$ evolves exactly as the parallel velocity of guiding center pseudo-particle.
The Lagrangian multipliers and the constraints contribute to the perpendicular velocity of pseudo-particle $\sum_{k=1}^{2}\lambda^{k}\nabla_p g^{k}$ and cause the drift motion.

%\begin{figure}
%\includegraphics[scale=0.8]{fig21}
%\caption{The orbits of a guiding center pseudo-particle (in red) in an external magnetic field and corresponding charged particle (in green) for reference. The direction of magnetic field (in blue), $\boldsymbol{p}-\boldsymbol{A}(\boldsymbol{X})$ (in pink), and perpendicular velocity of the guiding center (in black) are plotted at different positions along the trajectory. The initial value of charged particle is $\boldsymbol{x}_0=(0.3,0.2,-1.4)$ and $\boldsymbol{p}_0=()$}
%\label{fig1}
%\end{figure}

In Fig.\,\ref{fig2}, we deprive the guiding center dynamics of its intrinsic magnetic moment (IMM) and two primary constraints, for comparison, to demonstrate their contributions.
We calculate a guiding center trajectory in a magnetic mirror field, which is widely used to confine charged particles or plasmas for a variety of purposes \cite{gibson1960containment,anderson2023observation}.
The setup of the mirror field and other details of numerical calculations are introduced in the Supplemental Material \cite{wang2017accurate}.
When properly choosing its initial condition, the guiding center is bounced back by the magnetic mirror, with its reflection point denoted by a black solid circle.
If we drop both the IMM term in the canonical guiding center Hamiltonian in Eq.\,\eqref{eq:H1} and the two constraints in Eq.\,\eqref{eq:c3}, we obtain a charged-particle-like trajectory, depicted by the orange curve in Fig.\,\ref{fig2}.
Without the force by the intrinsic magnetic moment, the particle gradually deviates the guiding center orbits and finally runs out of the mirror field.
For the system with the complete Hamiltonian but without constraints, its trajectory approaches the guiding center orbit, but fast Larmor gyromotion still exists, see the light blue hollow circles in Fig.\,\ref{fig2}.
The complete canonical guiding center dynamics with constraints successfully eliminates the small timescale effects from the dynamics, see the red curve in Fig.\,\ref{fig2}

\begin{figure}
\includegraphics[scale=0.35]{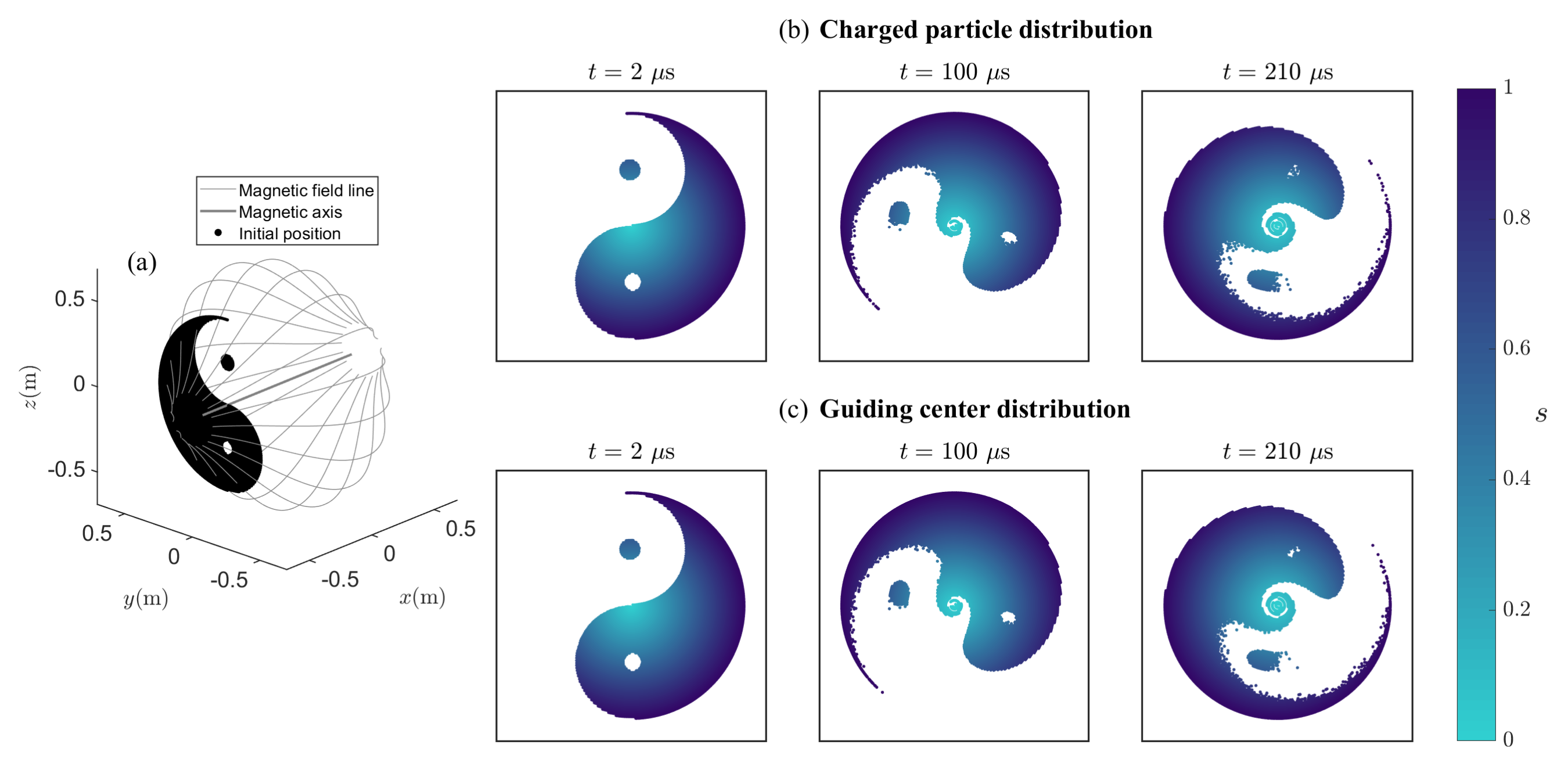}
\caption{(a) The initial 3D distribution of charged particles and guiding center particles in the magnetic mirror with electric field $E=(-200y/\sqrt{x^2+y^2},-200x/\sqrt{x^2+y^2},0)$. The spatial distributions projection in the y-z plane of charged particles and guiding center particles at three moments $t=2\mu s,~100\mu s,~210\mu s$ are plotted in subfigures (b) and (c), respectively. The color bar for $s$ indicates the corresponding normalized magnetic flux value.
}
\label{fig3}
\end{figure}

\begin{figure}
\includegraphics[scale=0.4]{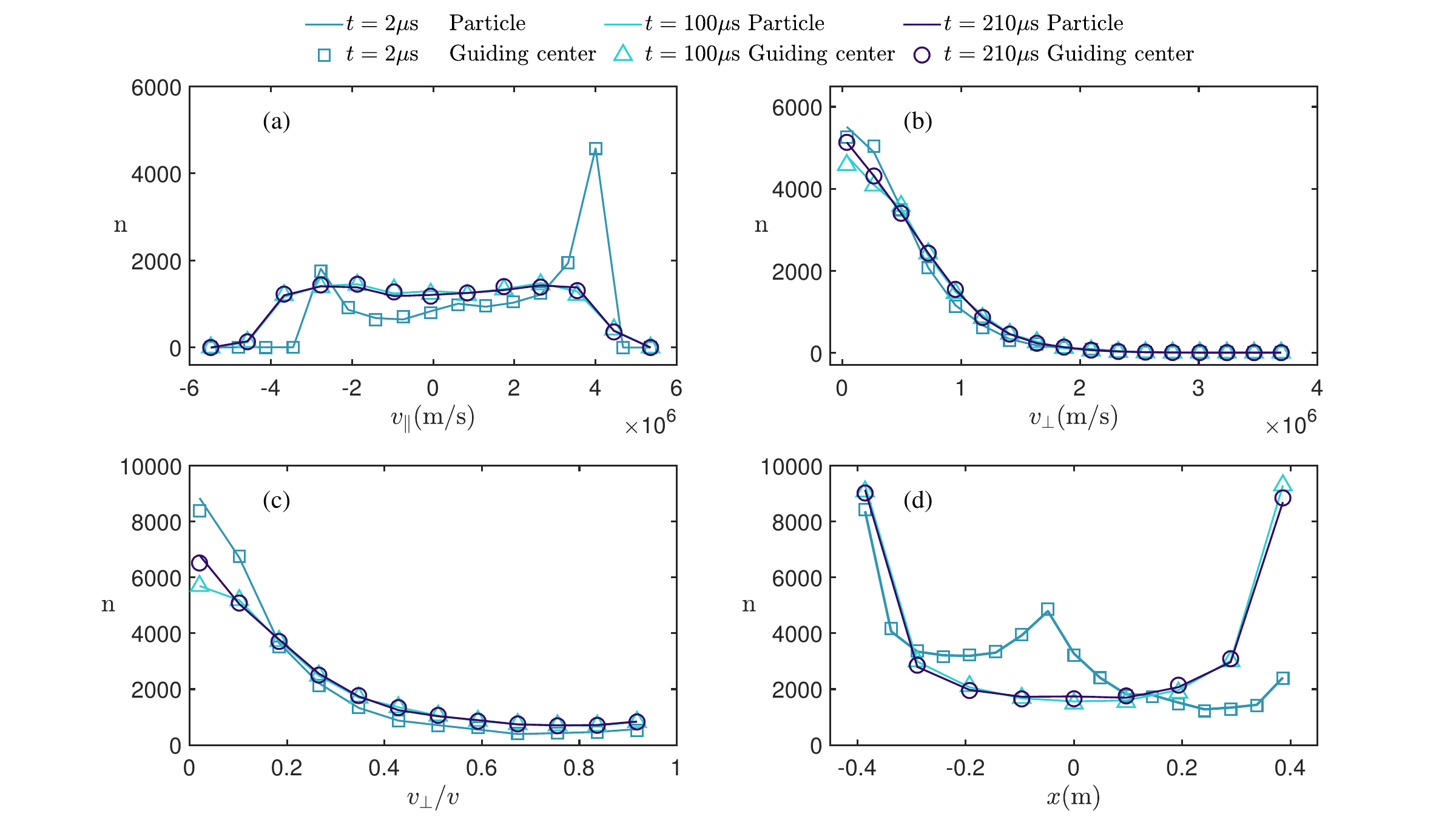}
\caption{The number distributions of the charged particles and the guiding center pseudo-particles with respect to pitch angle, parallel velocity, vertical velocity, and position in the $x$ direction at moments $t=2\mu s,~100\mu s,~210\mu s$. }
\label{fig4}
\end{figure}

To verify the canonical gyrokinetic equation in Eq.(\ref{eq:cgkinetic}), we calculate the evolution of distribution function in the magnetic mirror, using both the guiding center model and charged particle model.
The initial spatial distributions of guiding centers and particles are both set to a Tai Chi pattern in the y-z plane at  $x=-0.4$, see Fig.\,\ref{fig3} (a).
Due to the limitation of gyro-period, the time step for simulating charged particles is $\bigtriangleup t=5.882e^{-13} s$.
And the guiding center pseudo-particles are simulated using the time step $\bigtriangleup t=5.882e^{-12} s$, which is 10 times larger.
We plot the spatial distribution as well as corresponding normalized magnetic flux value at three moments $t=2\mu s,~100\mu s,~210\mu s$, respectively, in Fig.\,\ref{fig3} (b) and (c).
The spatial evolution of the two models are obviously consistent.
When transferring canonical guiding center distributions to corresponding particle representations, the number distribution with respect to pitch angle, parallel velocity, vertical velocity, and the position in the $x$ direction calculated by two different models are all compared and show strongly consistent, as depicted in Fig.\,\ref{fig4}.
In this case, the guiding center model describes the same physical distribution evolution as the charged particle model with only one-tenth of the time steps and therefore much less computational cost.

This letter introduces a comprehensive canonical Hamiltonian description of the guiding center dynamics.
Guiding center system can be generally expressed using canonical coordinates in a 6D phase space.
In this form, the guiding center is interpreted as a pseudo-particle with intrinsic magnetic moment and constraints independent of the particle motion.
The velocity and force of the pseudo-particle are determined by canonical dynamical equations of the pseudo-particle, where two primary constraints are key to eliminate fast-scale gyromotion.
Geometric structures, symplectic algorithms, and canonical gyrokinetic theories are readily developed based on the generalized canonicalization of guiding center systems.
In the future work, we will continue developing the theory of canonical gyro-kinetics and the canonical gyrokinetic PIC algorithms based on the canonical guiding center theory.
We will study the geometric structures and structure-preserving algorithms of guiding center system and their applications to real plasma problems.
Furthermore, this canonicalization scheme endows the pseudo particle with an intrinsic magnetic moment by restraining small scale dynamics, which is also an interesting topic and worth exploring.

\begin{acknowledgments}
This research is supported by  the National Natural Science Foundation of China (No. NSFC-12271025, NSFC-12171466), the State Key Laboratory of Alternate Electrical Power System with Renewable Energy Sources (Grant No. LAPS23003), and Geo-Algorithmic Plasma Simulator (GAPS) Project.

\end{acknowledgments}

%\bibliographystyle{apsrev4-1}
%\bibliography{Ref}


\end{document}